\documentclass[conference]{IEEEtran}
\IEEEoverridecommandlockouts
\usepackage{cite}
\usepackage{amsmath,amssymb,amsfonts}
\usepackage{algorithmic}
\usepackage{graphicx}
\usepackage{textcomp}

\usepackage{booktabs, multirow} 
\usepackage{soul}
\usepackage[table,xcdraw]{xcolor}
\usepackage{changepage,threeparttable} 
\usepackage{adjustbox}
\usepackage{verbatim}
\usepackage{tikz}

\def\BibTeX{{\rm B\kern-.05em{\sc i\kern-.025em b}\kern-.08em
    T\kern-.1667em\lower.7ex\hbox{E}\kern-.125emX}}

\usepackage{multirow}

\begin{document}

\newcommand\copyrighttext{%
  \footnotesize This paper is presented in IEEE Ural-Siberian Conference on Computational Technologies in Cognitive Science, Genomics and Biomedicine (CSGB)}
\newcommand\copyrightnotice{%
\begin{tikzpicture}[remember picture,overlay]
\node[anchor=south,yshift=10pt] at (current page.south) {\copyrighttext};
\end{tikzpicture}%
}

\newpage

\makeatletter 
\newcommand{\linebreakand}{%
  \end{@IEEEauthorhalign}
  \hfill\mbox{}\par
  \mbox{}\hfill\begin{@IEEEauthorhalign}
}
\makeatother 

\title{Benefits of mirror weight symmetry for 3D mesh segmentation in biomedical applications\\
\thanks{This work has been supported by the grant of the Russian Science Foundation, RSF 22-21-00930. The computations were performed on the Uran supercomputer at the IMM UB RAS.}
}

\author{
\IEEEauthorblockN{Vladislav Dordiuk}
\IEEEauthorblockA{\textit{Ural Federal University} \\
\textit{Institute of Immunology and Physiology}\\
Ekaterinburg, Russia \\
vladislav0860@gmail.com}

\and
\IEEEauthorblockN{Maksim Dzhigil}
\IEEEauthorblockA{\textit{Ural Federal University}\\
\textit{Institute of Immunology and Physiology}\\
Ekaterinburg, Russia \\
maxdzhigil@gmail.com}

\and
\IEEEauthorblockN{Konstantin Ushenin}
\IEEEauthorblockA{\textit{Institute of Natural Sciences} \\
\textit{Ural Federal University}\\
Ekaterinburg, Russia \\
konstantin.ushenin@urfu.ru}
}

\maketitle
\copyrightnotice

\begin{abstract} 
3D mesh segmentation is an important task with many biomedical applications. The human body has bilateral symmetry and some variations in organ positions. It allows us to expect a positive effect of rotation and inversion invariant layers in convolutional neural networks that perform biomedical segmentations. In this study, we show the impact of weight symmetry in neural networks that perform 3D mesh segmentation. We analyze the problem of 3D mesh segmentation for pathological vessel structures (aneurysms) and conventional anatomical structures (endocardium and epicardium of ventricles). Local geometrical features are encoded as sampling from the signed distance function, and the neural network performs prediction for each mesh node. We show that weight symmetry gains from 1 to 3\% of additional accuracy and allows decreasing the number of trainable parameters up to 8 times without suffering the performance loss if neural networks have at least three convolutional layers. This also works for very small training sets.

\end{abstract}

\begin{IEEEkeywords}
3D mesh segmentation, biomedical segmentation, weight symmetry, rotation invariant, inversion invariant, hard constraints, symmetry in neural networks
\end{IEEEkeywords}

\section{Introduction}

Convolutional neural networks successfully solve semantic and instance segmentation problems in biomedical applications. Results of neural network segmentation are especially notable in the segmentation of 3D imaging data such as computed tomography, magnetic resonance tomography \cite{khan2022sequential}, \textit{etc.} Segmentation of such type of data provides 3D voxel models of abdominal organs. However, the creation of voxel models is only a first step for modern biomedical pipelines. Medical visualization requires a surface mesh of objects. Anatomical measurements require key points. Production of personalized prosthetics requires surface mesh with advanced processing. Personalized computer simulations based on finite element analysis require a 3D mesh of volumetric elements \cite{salvador2021electromechanical, diaz2014computational}. 

All mentioned applications require segmentation of the surface mesh according to geometrical features or anatomical structures. 3D mesh segmentation is especially important for the detection of abnormal anatomical structures such as aneurysms \cite{intra3d}, and the segmentation of organs according to their conventional anatomical structures.

3D mesh segmentation in biomedical applications has some limitations and advantages in comparison with 3D mesh segmentation in computer graphics and computer vision. Medical datasets usually include smaller number of cases. That leads to significant issues with the lack of cases available in the training dataset. Segmentation of some geometrical features is invariant to inversions because the human body is bilaterally symmetrical. Some structures also are invariant to rotation because of slight variations in organ positions in the body. Thus, we expect a positive effect of rotation and inversion invariant layers on the performance of neural networks in 3d mesh segmentation tasks.

Some approaches to rotation invariant neural network are presented in \cite{alt2022designing, worrall2018cubenet, cabrera2017deep,dieleman2015rotation}. A wide range of methods for invariancy is described in \cite{geometry_book}. Unfortunately, most of the mentioned approaches significantly increase the number of trainable parameters. The other approach to achieve imperfect rotation invariancy is data augmentation. The augmentation pipeline does not enforce changes in neural network architecture, however it requires additional hyperparameter tuning for each specific task and increases the required computational resources.

In our study, we implement imperfect rotation and inversion symmetry of layers using mirror weight symmetry of convolutional neural networks \cite{deep_symmetry}. This approach cannot achieve mathematically correct invariancy to some transformations of the input. However, weight symmetry is straightforward and simple to implement. This approach does not increase the number of parameters in the neural network and does not require data augmentation. 

\begin{figure*}[hbt!] 
\begin{center}
\includegraphics[width=0.99\textwidth,  height=0.26\textwidth]{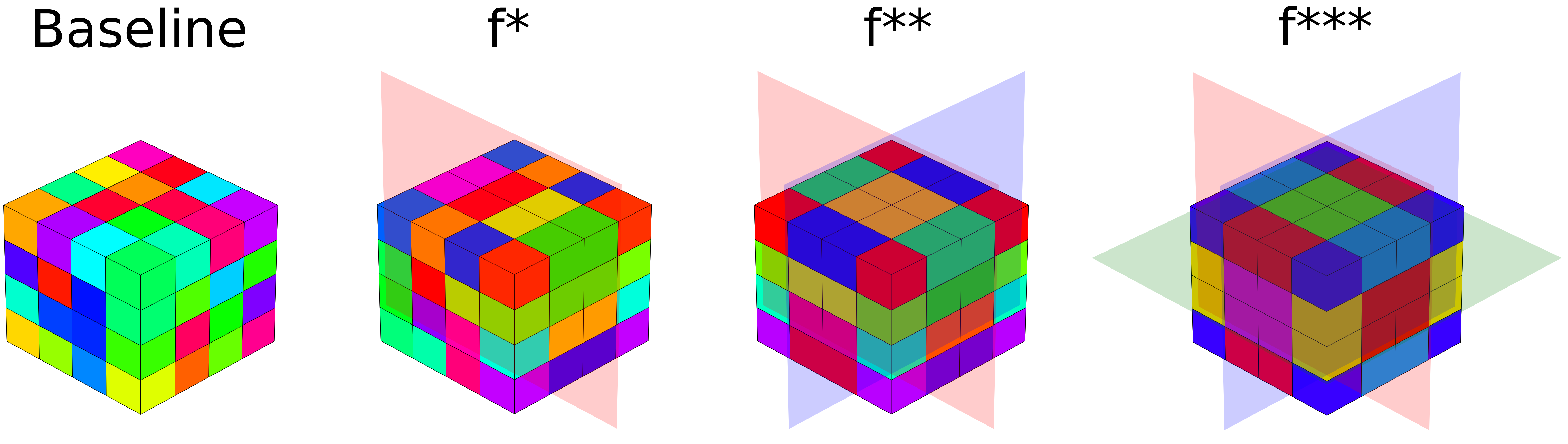}
\caption{Different types of symmetry, applied to a weight of a convolution layer with the kernel size of $4 \times 4 \times 4$. The intersecting planes show the axis along which symmetry can be observed.}\label{fig:symmetry}
\end{center}
\end{figure*}

To show the benefits of weight symmetry, we use small neural networks with 130,000 -- 700,000 parameters, which is 10-30 times less than the recent advanced approaches for 3D mesh segmentation \cite{intra3d}. Our neural networks predict the class of each node on the surface mesh according to the geometrical features of the local region. The local region is encoded with sampling from the signed distance function of the 3D object. Our study shows that this simple approach can segment 3D surface mesh with high precision, and can work with small train sets. Also, we show that weight symmetry in convolutional kernels improves the quality of segmentation and drastically reduces the number of the required parameters in the neural networks.

\section{Methods}

\subsection{Datasets}

In order to evaluate the performance of the neural networks and to study the effects of applying different types of weight symmetry, we use two datasets for 3D mesh segmentation. The first dataset is IntrA \cite{intra3d}, it includes 116 3D meshes of blood vessels with aneurysms, that were reconstructed from magnetic resonance angiogram images of anonymous patients. Every point of each mesh is marked with a label that classifies it as a healthy part of a blood vessel or as an aneurysm. 

In our work, we used the original labeling for two classes, but modified the surface of 3D models. We closed hollows on the sides of blood vessels using the "Close hole" function from Meshlab \cite{meshlab} software. This processing complements meshes to closed surfaces and makes it possible to compute the signed distance function from these surfaces.


The second dataset is derived from Automated Cardiac Diagnosis Challenge (ACDC) \cite{ACDC}. ACDC is a challenge for methods of magnetic resonance image segmentation. The training part of the ACDC dataset includes scans of ventricles for 100 patients. Labeling was performed by experts for the 3D volume region of the ventricular myocardium, left atrium cavity, and right atrium cavity.


We used the ground truth labels for the training dataset, provided for this challenge, in order to reconstruct 3D meshes of hearts. For that, we used marching cubes algorithm and Taubin smoothing. After the meshes were reconstructed, we labeled them, dividing each mesh into three classes: epicardium, left ventricular endocardium, and right ventricular endocardium. In order to avoid confusion, in the next sections we will refer to this dataset as ACDC-S, where S stands for surface, meaning that we work with a surface of reconstructed 3D mesh. 

In summary, we perform a study on two 3D mesh segmentation datasets. The first one (IntrA) contains two classes and a task to locate the aneurysm on a blood vessel. The second one (ACDC-S) contains three classes with a task to mark up epicardium, left ventricular endocardium, and right ventricular endocardium. We divide each dataset with $9$ different train-test ratios. They include $10:90$, $20:80$, $30:70$, $40:60$, $50:50$, $60:40$, $70:30$, $80:20$, and $90:10$ ratios, where the first number stands for the portion of data, that would go into the train split and the second number refers to the test split. So every neural network would be trained with different splits a total of $9$ times for each dataset. This approach to train-test splitting allows us to show that our methods are able to work with small train sets, which simulates the lack of data. This makes our study close to real biomedical applications, when number of available patients data is scarce.

\subsection{Local geometry encoding}

In our approach, the neural network predicts the class of each vertex of 3D mesh, based on their local geometry features. In order to implement this, we used the encoding with signed distance function (SDF) \cite{deepsdf}. It represents the geometry as a distance matrix, where the points outside an enclosed surface have positive distances and the distances of the points on the inside are negative. 

Our processing goes as follows. We define a neighborhood of a point (x,y,z) as a cube $[x-\frac{a}{2}; x+\frac{a}{2}] \times [y-\frac{a}{2}; y+\frac{a}{2}] \times [z-\frac{a}{2}; z+\frac{a}{2}]$ with $N$ points on each side. In this cube, we create a uniform mesh for data sampling. A point of a mesh is defined as $\overline{\boldsymbol{p}}_{ijk}$ $= \big((x -\frac{a}{2}) + ih; (x -\frac{a}{2}) + jh; (x -\frac{a}{2}) + kh\big)$, where $h= \frac{a}{N-1}$ is the spacing between the points in the cube. 

To segment the surface of 3D objects the neural network (NN) solves the problem of multi-label classification: $c = \operatorname{NN}\big(\{\{\{\operatorname{SDF}_r(\overline{\boldsymbol{p}}_{ijk})\}_{i=1}^{N}  \}_{j=1}^{N}   \}_{k=1}^{N}\big)$ where $\operatorname{SDF}_r(\cdot)$ is a signed distance to the boundary of the geometry model $r$, and $c$ is a class of point $(x,y,z)$ ($c\in [0, \boldsymbol{c}_{max}]$, $\boldsymbol{c}_{max} \in \mathbb{N}$). 

\subsection{Weight symmetry}

In this paper we work with 3D convolution layers, so we implemented three types of mirror symmetry. The first type is f*, it is a symmetry along one axis. Since we pass a 3D tensor as an input to the neural network, there are three possible symmetries: f1, f2, and f3. They correspond to x, y, and z axes in the original medical datasets that were discussed in the previous sections. The second type of symmetry is f**, that corresponds to the symmetry along two axes, this includes f12, f13 and f23 symmetries. Finally, f*** refers to the symmetry along all three axes and represents f123 symmetry. 

The Fig. \ref{fig:symmetry} shows how different symmetries affect a weight of 3D convolution layer with kernel size of $4 \times 4 \times 4$. We obtain the baseline values for each convolution kernel with Kaiming weight initialization \cite{Kaiming}, which is the default initialization method for PyTorch library \cite{paszke2019pytorch}. After this, mirror symmetries are applied to the weights using the flip of the original tensor and concatenation. 

The applying of symmetry along one axis could be described as $\boldsymbol{\hat{W}} = W \oplus T_a (W) $, where W is the weight of the convolution layer, $T_a$ is the flip, that reverses the order of elements in the tensor along the axis $a$, and $\oplus$ is concatenation along the same axis. The symmetry along several axes requires the repeat of the described transformation for every axis sequentially. 

If we assume that the baseline tensor contains ($N \times N \times N$) elements, then in order to preserve the original kernel size with the use of f* symmetry, we would have to halve the number of parameters along the symmetry axis. This is needed as the concatenation would double the size of the tensor. This means that for f1 symmetry, we would use a tensor of ($\frac{N}{2} \times N \times N$) size, since after applying the symmetry along the axis 1, it would have the needed shape of ($N \times N \times N$). For f2 the tensor would be ($N \times  \frac{N}{2} \times N$), and f3 symmetry would need ($N \times N \times \frac{N}{2} $) elements. 
The f** symmetry would require two concatenations, so we need to halve the tensor size along an additional axis. Thus, for f12 symmetry we would use weights with the size of ($\frac{N}{2} \times \frac{N}{2} \times N$), for f13 they would be ($\frac{N}{2} \times N \times \frac{N}{2} $), and f23 needs a tensor with ($N \times \frac{N}{2} \times \frac{N}{2}$) parameters.
The convolution with f*** symmetry would require halving the size of the baseline tensor along all axes. So, f123 would require a tensor with the shape of ($\frac{N}{2} \times \frac{N}{2} \times \frac{N}{2}$). 
This allows us to decrease the amount of unique trainable parameters in each layer by two times with f* symmetries, by four times with f** and by eight times with f***, while being able to use the same architecture of the neural network.

\subsection{Neural networks}

The use of symmetric weights reduces the amount of trainable parameters of neural networks, this leaves two experimental designs to consider. We can either keep the architecture and allow the number of parameters to significantly drop, or we can modify it to keep the number of parameters by increasing the number of convolution kernels. In this paper we study both approaches and refer to them as tasks.

\begin{figure}[h!] 
\begin{center}
\includegraphics[width=0.48\textwidth,  height=0.43\textwidth]{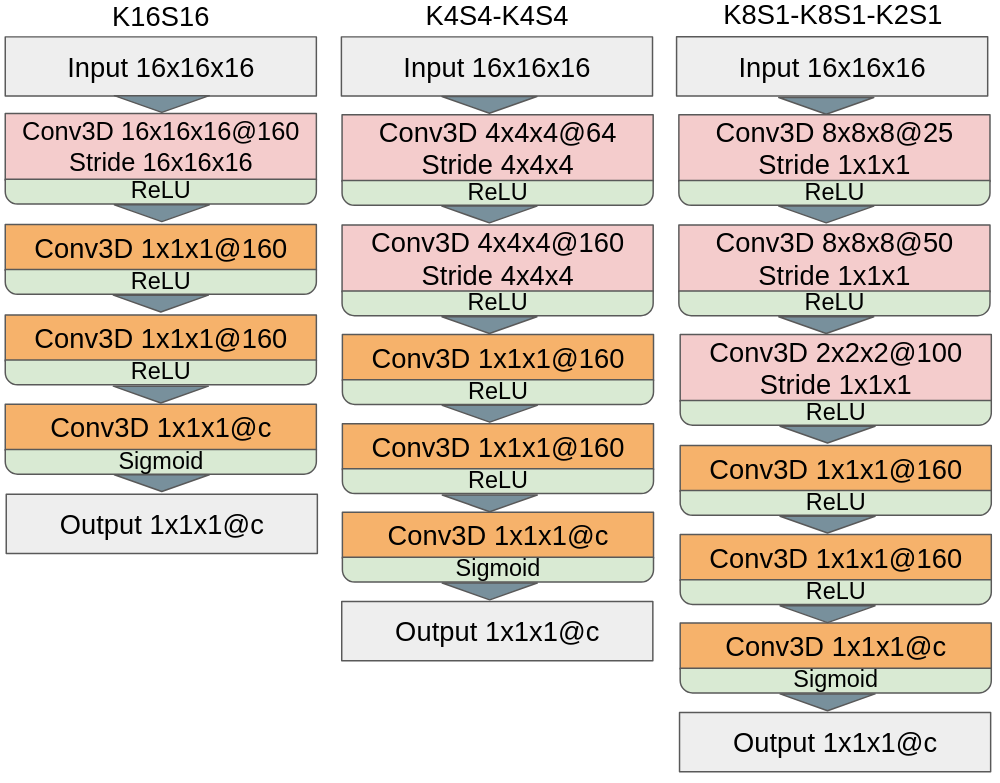}
\caption{Neural network architectures. $c$ in the last layer and output stands for the number of classes in the dataset.}\label{fig:model}
\end{center}
\end{figure}

\begin{table}[]
\caption{Number of Kernels That Were Used to Keep the Number of Trainable Parameters in the Models With Symmetric Convolution Weights. The MLP Block is Marked With Color.}\label{tab:kernels}
\centering
\begin{tabular}{ccccc}
\multicolumn{5}{c}{K16S16}                                                                 \\ \hline
\multicolumn{1}{c|}{}                              & \multicolumn{4}{c}{Number of kernels} \\ \cline{2-5} 
\multicolumn{1}{c|}{layers}                        & baseline    & f*     & f**   & f***   \\ \hline
\multicolumn{1}{c|}{16x16x16}                      & 160         & 308    & 575   & 1012   \\ \hline
\rowcolor[HTML]{FFCC67} 
\multicolumn{1}{c|}{\cellcolor[HTML]{FFCC67}1x1x1} & 160         & 160    & 160   & 160    \\
\rowcolor[HTML]{FFCC67} 
\multicolumn{1}{c|}{\cellcolor[HTML]{FFCC67}1x1x1} & 160         & 160    & 160   & 160    \\
\rowcolor[HTML]{FFCC67} 
\multicolumn{1}{c|}{\cellcolor[HTML]{FFCC67}1x1x1} & c           & c      & c     & c      \\
                                                   &             &        &       &        \\
                                                   &             &        &       &        \\
\multicolumn{5}{c}{K4S4-K4S4}                                                              \\ \hline
\multicolumn{1}{c|}{}                              & \multicolumn{4}{c}{Number of kernels} \\ \cline{2-5} 
\multicolumn{1}{c|}{layers}                        & baseline    & f*     & f**   & f***   \\ \hline
\multicolumn{1}{c|}{4x4x4}                         & 64          & 80     & 132   & 225    \\
\multicolumn{1}{c|}{4x4x4}                         & 160         & 251    & 300   & 348    \\ \hline
\rowcolor[HTML]{FFCC67} 
\multicolumn{1}{c|}{\cellcolor[HTML]{FFCC67}1x1x1} & 160         & 160    & 160   & 160    \\
\rowcolor[HTML]{FFCC67} 
\multicolumn{1}{c|}{\cellcolor[HTML]{FFCC67}1x1x1} & 160         & 160    & 160   & 160    \\
\rowcolor[HTML]{FFCC67} 
\multicolumn{1}{c|}{\cellcolor[HTML]{FFCC67}1x1x1} & c           & c      & c     & c      \\
                                                   &             &        &       &        \\
                                                   &             &        &       &        \\
\multicolumn{5}{c}{K8S1-K8S1-K2S1}                                                         \\ \hline
\multicolumn{1}{c|}{}                              & \multicolumn{4}{c}{Number of kernels} \\ \cline{2-5} 
\multicolumn{1}{c|}{layers}                        & baseline    & f*     & f**   & f***   \\ \hline
\multicolumn{1}{c|}{8x8x8}                         & 25          & 35     & 50    & 73     \\
\multicolumn{1}{c|}{8x8x8}                         & 50          & 71     & 100   & 140    \\
\multicolumn{1}{c|}{2x2x2}                         & 100         & 145    & 175   & 167    \\ \hline
\rowcolor[HTML]{FFCC67} 
\multicolumn{1}{c|}{\cellcolor[HTML]{FFCC67}1x1x1} & 160         & 160    & 160   & 160    \\
\rowcolor[HTML]{FFCC67} 
\multicolumn{1}{c|}{\cellcolor[HTML]{FFCC67}1x1x1} & 160         & 160    & 160   & 160    \\
\rowcolor[HTML]{FFCC67} 
\multicolumn{1}{c|}{\cellcolor[HTML]{FFCC67}1x1x1} & c           & c      & c     & c     
\end{tabular}
\end{table}

\begin{table*}[htp!]

\caption{Overview of Used Architectures. }

\begin{adjustbox}{width=1\textwidth}
\small

\begin{tabular}{cc|cccccccccccc}
\multicolumn{1}{c|}{Task}                          & Model          & base & f*  & f** & \multicolumn{1}{c|}{f***} & Mean         & \multicolumn{1}{c|}{Std.}       & base    & f*      & f**     & f***                         & Mean             & Std.         \\ \hline
\multicolumn{1}{c|}{Keep the number of parameters} & K16S16         & 160  & 308 & 575 & \multicolumn{1}{c|}{1012} & 514          & \multicolumn{1}{c|}{374}        & 707,362 & 706,614 & 707,617 & \multicolumn{1}{c|}{707,318} & \textbf{707,228} & \textbf{430} \\
\multicolumn{1}{c|}{}                              & K4S4-K4S4      & 224  & 331 & 432 & \multicolumn{1}{c|}{573}  & 390          & \multicolumn{1}{c|}{149}        & 711,522 & 711,853 & 710,386 & \multicolumn{1}{c|}{710,695} & \textbf{711,114} & \textbf{688} \\
\multicolumn{1}{c|}{}                              & K8S1-K8S1-K2S1 & 175  & 251 & 325 & \multicolumn{1}{c|}{380}  & 283          & \multicolumn{1}{c|}{89}         & 735,217 & 735,993 & 735,967 & \multicolumn{1}{c|}{735,474} & \textbf{735,663} & \textbf{381} \\ \hline
\multicolumn{1}{c|}{Keep the architecture}         & K16S16         & 160  & 160 & 160 & \multicolumn{1}{c|}{160}  & \textbf{160} & \multicolumn{1}{c|}{\textbf{0}} & 707,362 & 379,682 & 215,842 & \multicolumn{1}{c|}{133,922} & 359,202          & 253,599      \\
\multicolumn{1}{c|}{}                              & K4S4-K4S4      & 224  & 224 & 224 & \multicolumn{1}{c|}{224}  & \textbf{224} & \multicolumn{1}{c|}{\textbf{0}} & 711,522 & 381,794 & 216,930 & \multicolumn{1}{c|}{134,498} & 361,186          & 255,184      \\
\multicolumn{1}{c|}{}                              & K8S1-K8S1-K2S1 & 175  & 175 & 175 & \multicolumn{1}{c|}{175}  & \textbf{175} & \multicolumn{1}{c|}{\textbf{0}} & 735,217 & 388,817 & 215,617 & \multicolumn{1}{c|}{129,017} & 367,167          & 268,087      \\ \hline
                                                   &                & \multicolumn{6}{c|}{Number of convolution kernels}                                             & \multicolumn{6}{c}{Number of parameters}                                                    
\end{tabular}
\end{adjustbox}
\label{tab:arch}
\end{table*}

First, we made three fully convolutional neural network architectures to work as baselines. They are shown in Fig. \ref{fig:model}. Each model consists of one, two or three convolutional layers, followed by an MLP block. This block is the same for each architecture, and it includes two $1 \times 1 \times 1$ convolution layers with $160$ kernels each and an output layer with $c$ kernels, where $c$ is a number of classes in the dataset. All models use ReLU activation in deep layers and Sigmoid in the output layer. To distinguish the networks, we give each of them a name that reflects their number of layers, the shape of convolution and its stride. For instance, the notation K4S4-K4S4 means that the neural network includes two 3D convolution layers with kernel size and stride of $4 \times 4 \times 4$.

The first task is to keep the number of trainable parameters in the models with symmetry the same as in the baselines. For each of the baseline architectures, we created 7 additional models. They include three options with f* symmetry, one for each axis, three models with f** symmetry, and one model with f*** symmetry. After that, we increased the number of kernels in every model with symmetry. The resulting architectures are described in the Table \ref{tab:kernels}, and the total number of kernels and parameters for each architecture is shown in Table \ref{tab:arch}. 

For the task of keeping the architecture, we used the same models described above, but kept the amount of convolution kernels the same as in the baselines. This resulted in significant loss of trainable parameters, that can be seen in the Table \ref{tab:arch}. 

Finally, we had 15 models for each of three architectures. This includes the baseline, 7 models with the same number of trainable parameters, and 7 models with the same number of convolution kernels. Training of 45 models with 9 different train-test splits on two datasets resulted in a total of 810 experiments, that were conducted on the Uran supercomputer with NVIDIA K40m GPUs.

\section{Results}

\begin{table*}[htp!]
\caption{Performance of the Models on Train-Test Splits. The Best Values for Each Task and Model are in Bold. The Numbers In Parentheses are the Difference Comparing to the Baseline.}
\label{tab:results}
\begin{adjustbox}{width=1\textwidth}
\small
\begin{tabular}{cccccccccc}
\multicolumn{10}{c}{\textbf{20 : 80 train-test split}} \\ \hline
\multicolumn{10}{c}{IntrA dataset} \\ \hline
\multicolumn{1}{c|}{Task} & \multicolumn{1}{c|}{Model} & baseline & f1 & f2 & f3 & f12 & f13 & f23 & f123 \\ \hline
\multicolumn{1}{c|}{\multirow{3}{*}{Keep the number of parameters}} & \multicolumn{1}{c|}{K16S16} & \textbf{78.70} & 75.93 (-2.77) & 76.28 (-2.42) & 76.43 (-2.27) & 74.99 (-3.71) & 74.06 (-4.64) & 74.56 (-4.14) & 73.87 (-4.83) \\
\multicolumn{1}{c|}{} & \multicolumn{1}{c|}{K4S4-K4S4} & 80.21 & 81.7 (+1.49) & 82.14 (+1.93) & 81.59 (+1.38) & 82.46 (+2.25) & 82.07 (+1.86) & \textbf{82.97 (+2.76)} & 82.36 (+2.15) \\
\multicolumn{1}{c|}{} & \multicolumn{1}{c|}{K8S1-K8S1-K2S1} & 80.99 & 81.67 (+0.68) & 81.75 (+0.76) & 81.4 (+0.41) & \textbf{83.57 (+2.58)} & 82.16 (+1.17) & 83.44 (+2.45) & 83.13 (+2.14) \\ \hline
\multicolumn{1}{c|}{\multirow{3}{*}{Keep the architecture}} & \multicolumn{1}{c|}{K16S16} & \textbf{78.70} & 75.94 (-2.76) & 76.3 (-2.4) & 76.14 (-2.56) & 74.25 (-4.45) & 74.26 (-4.44) & 74.19 (-4.51) & 73.46 (-5.24) \\
\multicolumn{1}{c|}{} & \multicolumn{1}{c|}{K4S4-K4S4} & 80.21 & 81.13 (+0.92) & 81.67 (+1.46) & 80.94 (+0.73) & 81.92 (+1.71) & 82.14 (+1.93) & \textbf{82.71 (+2.5)} & 82.5 (+2.29) \\
\multicolumn{1}{c|}{} & \multicolumn{1}{c|}{K8S1-K8S1-K2S1} & 80.99 & 81.26 (+0.27) & 81.2 (+0.21) & 81.52 (+0.53) & \textbf{83.24 (+2.25)} & 82.01 (+1.02) & 82.72 (+1.73) & 82.29 (+1.3) \\ \hline
\multicolumn{10}{c}{ACDC-S dataset} \\ \hline
\multicolumn{1}{c|}{Task} & \multicolumn{1}{c|}{Model} & baseline & f1 & f2 & f3 & f12 & f13 & f23 & f123 \\ \hline
\multicolumn{1}{c|}{\multirow{3}{*}{Keep the number of parameters}} & \multicolumn{1}{c|}{K16S16} & 88.09 & 85.19 (-2.9) & 87.7 (-0.39) & \textbf{88.24 (+0.15)} & 75.52 (-12.57) & 83.52 (-4.57) & 83.91 (-4.18) & 68.38 (-19.71) \\
\multicolumn{1}{c|}{} & \multicolumn{1}{c|}{K4S4-K4S4} & 84.58 & 82.06 (-2.52) & 85.89 (+1.31) & 84.72 (+0.14) & 81.15 (-3.43) & 82.13 (-2.45) & \textbf{87.35 (+2.77)} & 78.89 (-5.69) \\
\multicolumn{1}{c|}{} & \multicolumn{1}{c|}{K8S1-K8S1-K2S1} & \textbf{90.05} & 88.02 (-2.03) & 90.02 (-0.03) & 88.35 (-1.7) & 89.92 (-0.13) & 87.89 (-2.16) & 90.03 (-0.02) & 89.85 (-0.2) \\ \hline
\multicolumn{1}{c|}{\multirow{3}{*}{Keep the architecture}} & \multicolumn{1}{c|}{K16S16} & 88.09 & 85.57 (-2.52) & 87.47 (-0.62) & \textbf{88.27 (+0.18)} & 77.5 (-10.59) & 83.56 (-4.53) & 84.78 (-3.31) & 67.97 (-20.12) \\
\multicolumn{1}{c|}{} & \multicolumn{1}{c|}{K4S4-K4S4} & 84.58 & 83.94 (-0.64) & \textbf{87.48 (+2.9)} & 84.21 (-0.37) & 77.63 (-6.95) & 82.56 (-2.02) & 85.99 (+1.41) & 79.34 (-5.24) \\
\multicolumn{1}{c|}{} & \multicolumn{1}{c|}{K8S1-K8S1-K2S1} & 90.05 & 88.32 (-1.73) & 89.99 (-0.06) & 89.46 (-0.59) & 90.01 (-0.04) & 87.4 (-2.65) & \textbf{90.84 (+0.79)} & 88.98 (-1.07) \\ \hline
\multicolumn{1}{l}{} & \multicolumn{1}{l}{} & \multicolumn{1}{l}{} & \multicolumn{1}{l}{} & \multicolumn{1}{l}{} & \multicolumn{1}{l}{} & \multicolumn{1}{l}{} & \multicolumn{1}{l}{} & \multicolumn{1}{l}{} & \multicolumn{1}{l}{} \\
\multicolumn{10}{c}{\textbf{50 : 50 train-test split}} \\ \hline
\multicolumn{10}{c}{IntrA dataset} \\ \hline
\multicolumn{1}{c|}{Task} & \multicolumn{1}{c|}{Model} & baseline & f1 & f2 & f3 & f12 & f13 & f23 & f123 \\ \hline
\multicolumn{1}{c|}{\multirow{3}{*}{Keep the number of parameters}} & \multicolumn{1}{c|}{K16S16} & \textbf{81.09} & 78.83 (-2.26) & 78.18 (-2.91) & 77.93 (-3.16) & 76.8 (-4.29) & 76.4 (-4.69) & 76.11 (-4.98) & 74.57 (-6.52) \\
\multicolumn{1}{c|}{} & \multicolumn{1}{c|}{K4S4-K4S4} & 83.77 & \textbf{84.5 (+0.73)} & 83.16 (-0.61) & 83.48 (-0.29) & 84.28 (+0.51) & 84.19 (+0.42) & 83.96 (+0.19) & 83.92 (+0.15) \\
\multicolumn{1}{c|}{} & \multicolumn{1}{c|}{K8S1-K8S1-K2S1} & 82.63 & 84.39 (+1.76) & 84.84 (+2.21) & 84.65 (+2.02) & 85.41 (+2.78) & 85.67 (+3.04) & 85.41 (+2.78) & \textbf{85.73 (+3.1)} \\ \hline
\multicolumn{1}{c|}{\multirow{3}{*}{Keep the architecture}} & \multicolumn{1}{c|}{K16S16} & \textbf{81.09} & 78.95 (-2.14) & 78.12 (-2.97) & 78.53 (-2.56) & 76.13 (-4.96) & 76.31 (-4.78) & 75.79 (-5.3) & 74.43 (-6.66) \\
\multicolumn{1}{c|}{} & \multicolumn{1}{c|}{K4S4-K4S4} & 83.77 & 84.06 (+0.29) & 83.15 (-0.62) & 83.51 (-0.26) & 83.44 (-0.33) & \textbf{84.18 (+0.41)} & 83.84 (+0.07) & 83.85 (+0.08) \\
\multicolumn{1}{c|}{} & \multicolumn{1}{c|}{K8S1-K8S1-K2S1} & 82.63 & 84.16 (+1.53) & 84.53 (+1.9) & 84.43 (+1.8) & 84.68 (+2.05) & 85.44 (+2.81) & 85.23 (+2.6) & \textbf{85.77 (+3.14)} \\ \hline
\multicolumn{10}{c}{ACDC-S dataset} \\ \hline
\multicolumn{1}{c|}{Task} & \multicolumn{1}{c|}{Model} & baseline & f1 & f2 & f3 & f12 & f13 & f23 & f123 \\ \hline
\multicolumn{1}{c|}{\multirow{3}{*}{Keep the number of parameters}} & \multicolumn{1}{c|}{K16S16} & \textbf{92.68} & 88.4 (-4.28) & 91.22 (-1.46) & 90.59 (-2.09) & 81.73 (-10.95) & 86.64 (-6.04) & 88.4 (-4.28) & 70.34 (-22.34) \\
\multicolumn{1}{c|}{} & \multicolumn{1}{c|}{K4S4-K4S4} & 89.80 & 86.5 (-3.3) & 90.47 (+0.67) & 88.46 (-1.34) & 83.99 (-5.81) & 86.74 (-3.06) & \textbf{90.92 (+1.12)} & 82.95 (-6.85) \\
\multicolumn{1}{c|}{} & \multicolumn{1}{c|}{K8S1-K8S1-K2S1} & 92.22 & 92.06 (-0.16) & 92.4 (+0.18) & 92.52 (+0.3) & 92.34 (+0.12) & 91.3 (-0.92) & \textbf{92.72 (+0.5)} & 91.26 (-0.96) \\ \hline
\multicolumn{1}{c|}{\multirow{3}{*}{Keep the architecture}} & \multicolumn{1}{c|}{K16S16} & \textbf{92.68} & 89.21 (-3.47) & 91.31 (-1.37) & 91.24 (-1.44) & 81.38 (-11.3) & 86.73 (-5.95) & 89.01 (-3.67) & 72.47 (-20.21) \\
\multicolumn{1}{c|}{} & \multicolumn{1}{c|}{K4S4-K4S4} & \textbf{89.80} & 87.22 (-2.58) & 89.06 (-0.74) & 89.79 (-0.01) & 82.49 (-7.31) & 86.35 (-3.45) & 88.57 (-1.23) & 83.06 (-6.74) \\
\multicolumn{1}{c|}{} & \multicolumn{1}{c|}{K8S1-K8S1-K2S1} & 92.22 & 91.67 (-0.55) & \textbf{93.39 (+1.17)} & 92.79 (+0.57) & 92.21 (-0.01) & 92.18 (-0.04) & 93.2 (+0.98) & 93.35 (+1.13) \\ \hline
\multicolumn{1}{l}{} & \multicolumn{1}{l}{} & \multicolumn{1}{l}{} & \multicolumn{1}{l}{} & \multicolumn{1}{l}{} & \multicolumn{1}{l}{} & \multicolumn{1}{l}{} & \multicolumn{1}{l}{} & \multicolumn{1}{l}{} & \multicolumn{1}{l}{} \\
\multicolumn{10}{c}{\textbf{80 : 20 train-test split}} \\ \hline
\multicolumn{10}{c}{IntrA dataset} \\ \hline
\multicolumn{1}{c|}{Task} & Model & baseline & f1 & f2 & f3 & f12 & f13 & f23 & f123 \\ \hline
\multicolumn{1}{c|}{\multirow{3}{*}{Keep the number of parameters}} & \multicolumn{1}{c|}{K16S16} & \textbf{81.86} & 78.37 (-3.49) & 77.35 (-4.51) & 78.15 (-3.71) & 76.11 (-5.75) & 75.47 (-6.39) & 75.13 (-6.73) & 74.01 (-7.85) \\
\multicolumn{1}{c|}{} & \multicolumn{1}{c|}{K4S4-K4S4} & \textbf{85.20} & 85.09 (-0.11) & 84.39 (-0.81) & 84.25 (-0.95) & 84.9 (-0.3) & 84.57 (-0.63) & 84.31 (-0.89) & 84.17 (-1.03) \\
\multicolumn{1}{c|}{} & \multicolumn{1}{c|}{K8S1-K8S1-K2S1} & 85.03 & 85.57 (+0.54) & 85.14 (+0.11) & 85.51 (+0.48) & 85.54 (+0.51) & 85.24 (+0.21) & 85.18 (+0.15) & \textbf{85.64 (+0.61)} \\ \hline
\multicolumn{1}{c|}{\multirow{3}{*}{Keep the architecture}} & \multicolumn{1}{c|}{K16S16} & \textbf{81.86} & 77.91 (-3.95) & 78.3 (-3.56) & 78.39 (-3.47) & 76.15 (-5.71) & 75.29 (-6.57) & 75.26 (-6.6) & 73.28 (-8.58) \\
\multicolumn{1}{c|}{} & \multicolumn{1}{c|}{K4S4-K4S4} & \textbf{85.20} & 84.76 (-0.44) & 84.29 (-0.91) & 84.24 (-0.96) & 84.06 (-1.14) & 84.75 (-0.45) & 84.16 (-1.04) & 84.2 (-1.0) \\
\multicolumn{1}{c|}{} & \multicolumn{1}{c|}{K8S1-K8S1-K2S1} & 85.03 & 84.74 (-0.29) & 84.91 (-0.12) & 85.06 (+0.03) & 85.27 (+0.24) & 85.4 (+0.37) & 85.97 (+0.94) & \textbf{86.3 (+1.27)} \\ \hline
\multicolumn{10}{c}{ACDC-S dataset} \\ \hline
\multicolumn{1}{c|}{Task} & \multicolumn{1}{c|}{Model} & baseline & f1 & f2 & f3 & f12 & f13 & f23 & f123 \\ \hline
\multicolumn{1}{c|}{\multirow{3}{*}{Keep the number of parameters}} & \multicolumn{1}{c|}{K16S16} & \textbf{95.42} & 91.87 (-3.55) & 91.83 (-3.59) & 91.96 (-3.46) & 84.42 (-11.0) & 87.41 (-8.01) & 89.91 (-5.51) & 75.34 (-20.08) \\
\multicolumn{1}{c|}{} & \multicolumn{1}{c|}{K4S4-K4S4} & 91.52 & 89.23 (-2.29) & \textbf{92.0 (+0.48)} & 91.02 (-0.5) & 86.52 (-5.0) & 87.91 (-3.61) & 91.67 (+0.15) & 86.62 (-4.9) \\
\multicolumn{1}{c|}{} & \multicolumn{1}{c|}{K8S1-K8S1-K2S1} & \textbf{94.67} & 92.95 (-1.72) & 94.41 (-0.26) & 94.36 (-0.31) & 93.76 (-0.91) & 93.62 (-1.05) & 94.64 (-0.03) & 93.83 (-0.84) \\ \hline
\multicolumn{1}{c|}{\multirow{3}{*}{Keep the architecture}} & \multicolumn{1}{c|}{K16S16} & \textbf{95.42} & 92.19 (-3.23) & 93.02 (-2.4) & 93.09 (-2.33) & 85.31 (-10.11) & 88.64 (-6.78) & 90.57 (-4.85) & 73.3 (-22.12) \\
\multicolumn{1}{c|}{} & \multicolumn{1}{c|}{K4S4-K4S4} & 91.52 & 89.01 (-2.51) & 92.22 (+0.7) & \textbf{92.13 (+0.61)} & 85.8 (-5.72) & 88.42 (-3.1) & 90.75 (-0.77) & 86.4 (-5.12) \\
\multicolumn{1}{c|}{} & \multicolumn{1}{c|}{K8S1-K8S1-K2S1} & 94.67 & 93.64 (-1.03) & 94.57 (-0.1) & 93.55 (-1.12) & 92.94 (-1.73) & 93.64 (-1.03) & 94.42 (-0.25) & \textbf{94.89 (+0.22)} \\ \hline
\end{tabular}
\end{adjustbox}
\end{table*}

\begin{figure*}[htp!] 
\begin{center}
\includegraphics[width=0.9\textwidth,  height=1.3\textwidth]{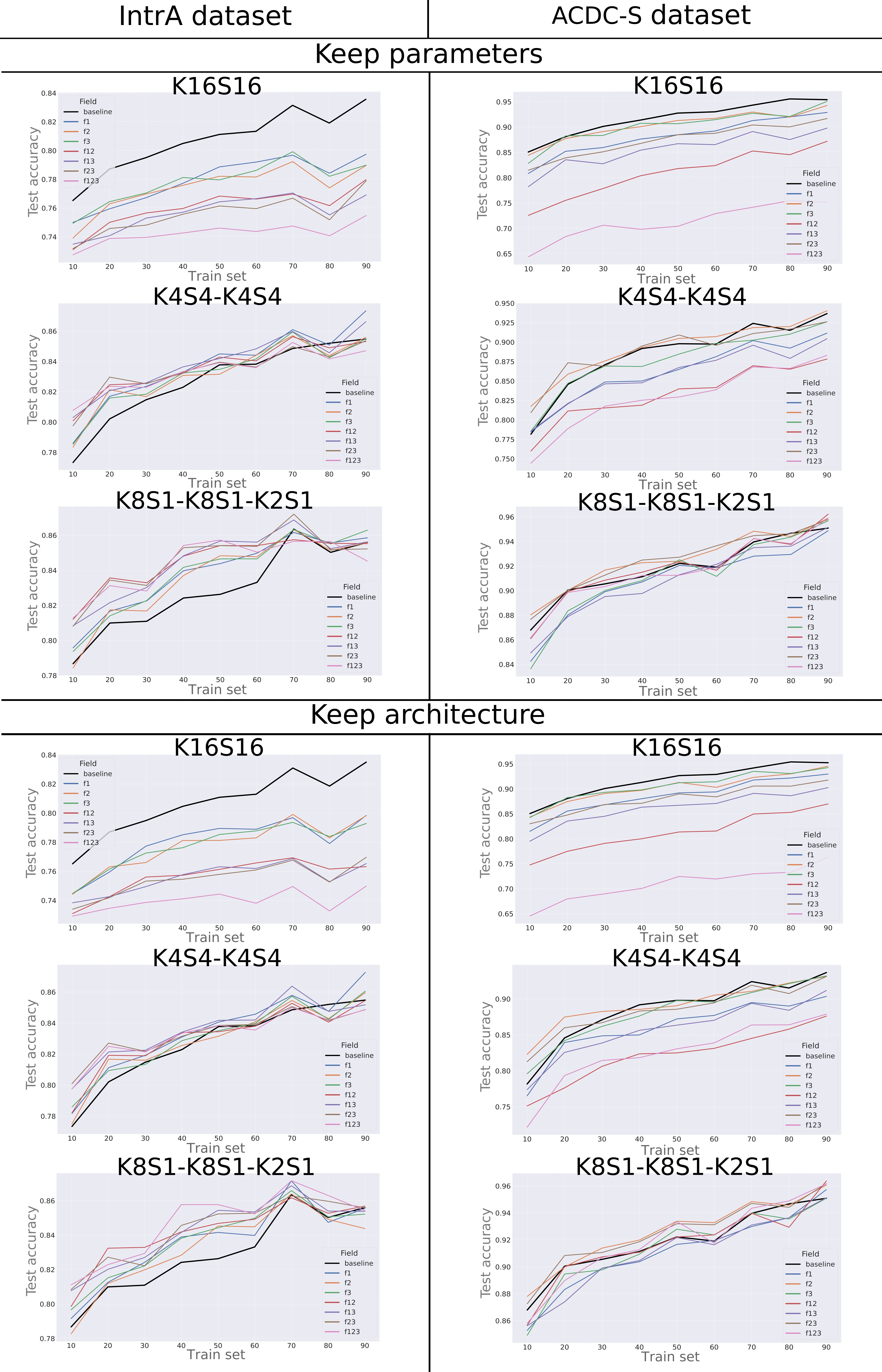}
\caption{Performance of different models on 9 train-test splits. The number after the letter “f” refers to the axes along which the symmetry is created.}\label{fig:graphs}
\end{center}
\end{figure*}

\begin{figure*}[htp!] 
\begin{center}
\includegraphics[width=1.0\textwidth,  height=1.1\textwidth]{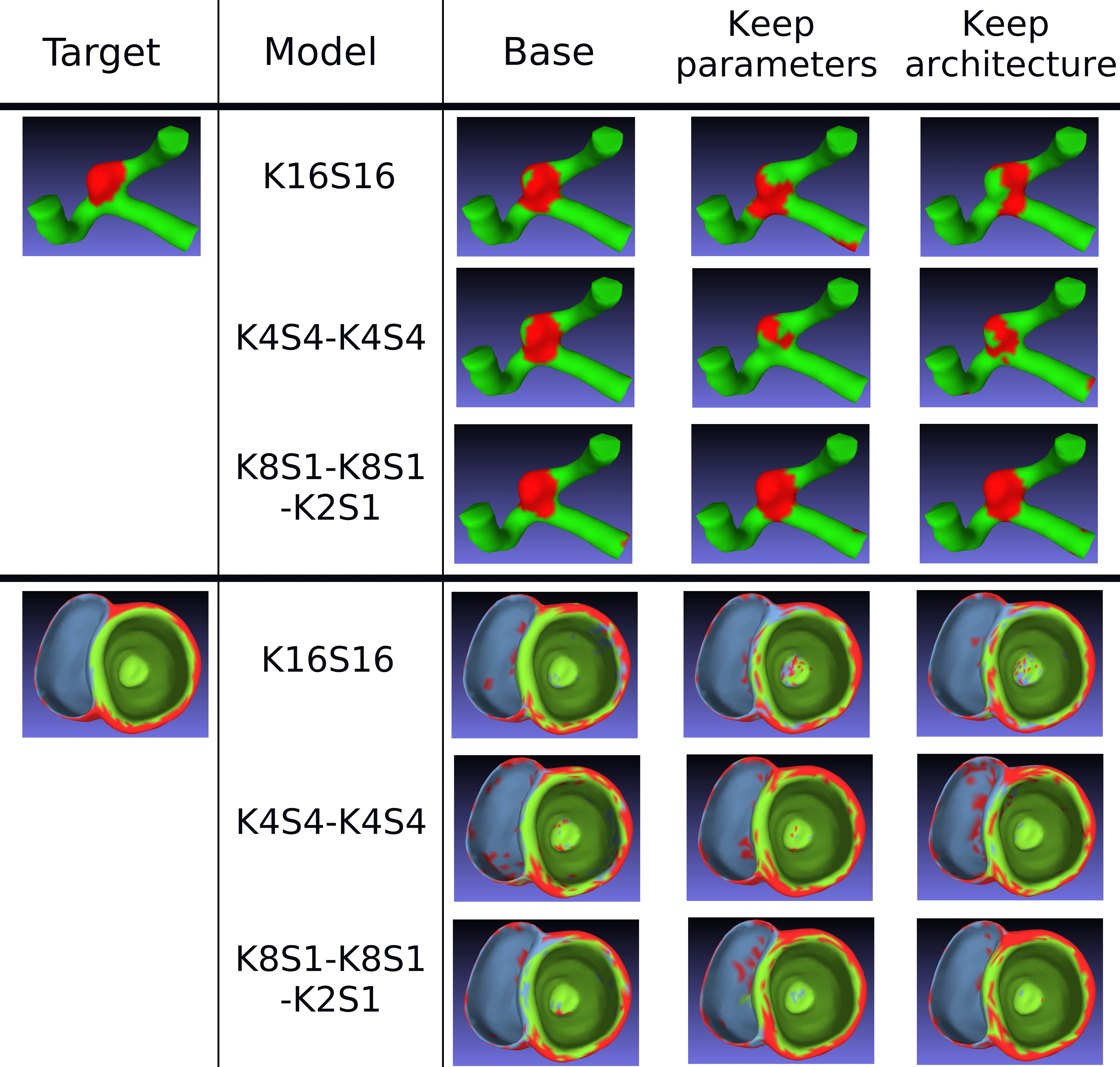}
\caption{Results of segmentation done by best performing models, trained on 20\% of dataset. Images on top are the models from IntrA dataset, the ones on the bottom are from ACDC-S.}\label{fig:meshes}
\end{center}
\end{figure*}

\subsection{Fixed number of parameters}
We compared the baselines to the models with symmetric layers, that have the same number of trainable parameters. The results show that the use of symmetric weights affects the performance, depending on the depth of the neural network, so that the deeper the network is, the more it would benefit from symmetry. Figure \ref{fig:graphs} and Table \ref{tab:results} demonstrate that the use of symmetry does not improve the performance of any model of K16S16 architecture, since it has only one convolution layer before the MLP block. The increase of convolutions up to two allows K4S4-K4S4 models with symmetry to reach and slightly outperform the baseline on IntrA dataset, and achieve the performance close to the baseline on ACDC-S dataset. K8S1-K8S1-K2S1 architecture shows notable improvement in the accuracy of the models with the use of symmetrical weights, as almost all models with symmetry outperform the baseline with every train-test split on IntrA, and show similar to the baseline results on ACDC-S.

\subsection{Fixed architecture}
In order to see how a model with fewer parameters, but with the same architecture would compare to the baseline, we kept the baseline architectures shown in Fig.\ref{fig:model} and applied symmetries to them without increasing the number of convolution kernels. The comparison revealed that the results of segmentation were almost identical to the ones described above. 
K16S16 models with symmetry showed a notable decrease in performance comparing to the baseline on both datasets, and the results were improving for the deeper networks. 
All K4S4-K4S4 models with symmetry achieved similar to the baseline performance on IntrA dataset, and most of them showed the results close to the baseline values on ACDC-S. 
As for the K8S1-K8S1-K2S1 architecture, its models with symmetry notably outperformed the baseline on IntrA dataset and showed the similar to the baseline results on ACDC-S.

\subsection{Comparison of the results}

In this study we tracked the change of performance among the models, that were trained with different train-test splits. The results have shown that the baseline accuracy decreases up to 8.14\% on IntrA dataset and up to 15.5\% on ACDC-S, as the amount of data in train set goes down from $90\%$ of the dataset to $10\%$. 
The Fig. \ref{fig:graphs} shows that the baseline performance on IntrA dataset ranges from $76.52\%$ to $83.5\%$ accuracy for K16S16 model, from $77.34\%$ to $85.48\%$ for K4S4-K4S4, and from $78.67\%$ to $85.6\%$ for K8S1-K8S1-K2S1. As for the ACDC-S dataset, K16S16 model shows $85.07\% - 95.26\%$ accuracy, K4S4-K4S4 shows the results of $78.17\% - 93.67\%$, and K8S1-K8S1-K2S1 leads to $86.79\% - 95.09\%$ accuracy. 
As we can see, the accuracy of mesh segmentation is almost $10\%$ higher on ACDC-S dataset comparing to IntrA. This could be possibly explained by the shape of the models in the datasets, since the heart meshes in ACDC-S are much more similar to each other than the blood vessels in IntrA.

The baseline values allowed us to evaluate the effect of the utilization of symmetric layers with different train-test splits. The Table \ref{tab:results} shows the change of accuracy for different types of symmetry for each model and dataset with $20:80$, $50:50$ and $80:20$ data split ratios. The best overall performance for each split was achieved by K8S1-K8S1-K2S1 model with some type of symmetry. For IntrA dataset, the best performance for these three splits was achieved by a model with f23 symmetry and two models with f123 symmetry, resulting in $83.57\%$, $85.77\%$, and $86.3\%$ accuracies respectively. For ACDC-S dataset the best models used f23, f2 and f123 symmetries, and showed $90.84\%$, $93.39\%$, and $94.89\%$ accuracy correspondingly.


We mentioned above that the models with fixed number of parameters and the ones with fixed architecture have shown similar results. Despite the accuracy values being very close, there were noticed some differences in their performance. In Table \ref{tab:results} we can see that with $20:80$ train-test split, most models with the fixed number of parameters show small increase in performance on both datasets comparing to the ones with fixed architectures.
The same behavior could be observed with $50:50$ train-test split, however the gap in performance is smaller. Finally, as the train set increases up to $80\%$ of the dataset, the difference between them becomes negligible.

The obtained findings suggest, that the increase in performance of models with symmetric weights is more dependent on the number of convolution layers than on the total number of trainable parameters. The Figure \ref{fig:meshes} confirms these results, as the deeper the network becomes, the fewer misclassifications are made with the models, that have symmetric weights. 
We can see that if the neural networks have less than three convolution layers, they struggle to identify the boundaries of aneurysm in the models from IntrA dataset, and label it only partly. At the same time, the model with three convolution layers was able to show results that were significantly more comparable to the baseline, covering the whole part of the blood vessel that includes the aneurysm. 
Similar results can be seen in ACDC-S dataset segmentation. It is clear that the baseline performance of K16S16 and K4S4-K4S4 models is better than the accuracy of the same models with symmetry. However, the model K8S1-K8S1-K2S1 that has three convolution layers produces results that are comparable to or better than the baseline.

\section{Conclusion}

Utilizing weight symmetry significantly reduces the number of trainable parameters in each convolution kernel. Here, we analyze two strategies to introduce weight symmetry in neural networks. The first approach is to keep the number of parameters with an increasing number of convolutional kernels. The second approach is to keep a number of convolutional kernels and significantly decrease the number of trainable parameters.

Both studied strategies of weight symmetry are suitable for some goals. The first strategy increases accuracy while maintaining the same number of parameters. The second strategy leads to a reduction in the number of trainable parameters in symmetric layers up to 8 times, with either a gain in performance or negligible decrease in accuracy.
We observe that the benefits of weight symmetry are notable only if the neural network architecture has at least three convolutional layers before the multi-layer perceptron. Weight symmetry provides an additional accuracy up to 3\% on small datasets. This effect is especially visible for small train sets, as the biggest improvement can be seen with train-test split with ratios from $10\% : 90\%$ to $50\% : 50\%$. The benefits of weight symmetry are not as notable on big train sets and advance the accuracy by only 1\%. It is notable, that our solution can segment the IntraA dataset with almost the same quality that is presented in \cite{intra3d} and uses 10-30 times fewer parameters.

\bibliographystyle{IEEEtran}
\bibliography{bibliography}

\end{document}